\documentclass[pre,twocolumn,showpacs,floatfix,amsmath,amssymb]{revtex4-1}

\usepackage{graphicx,color}
\usepackage{verbatim}

\newcommand{\figOne}{
\begin{figure}[t]
\centering
\includegraphics[height=1.25in]{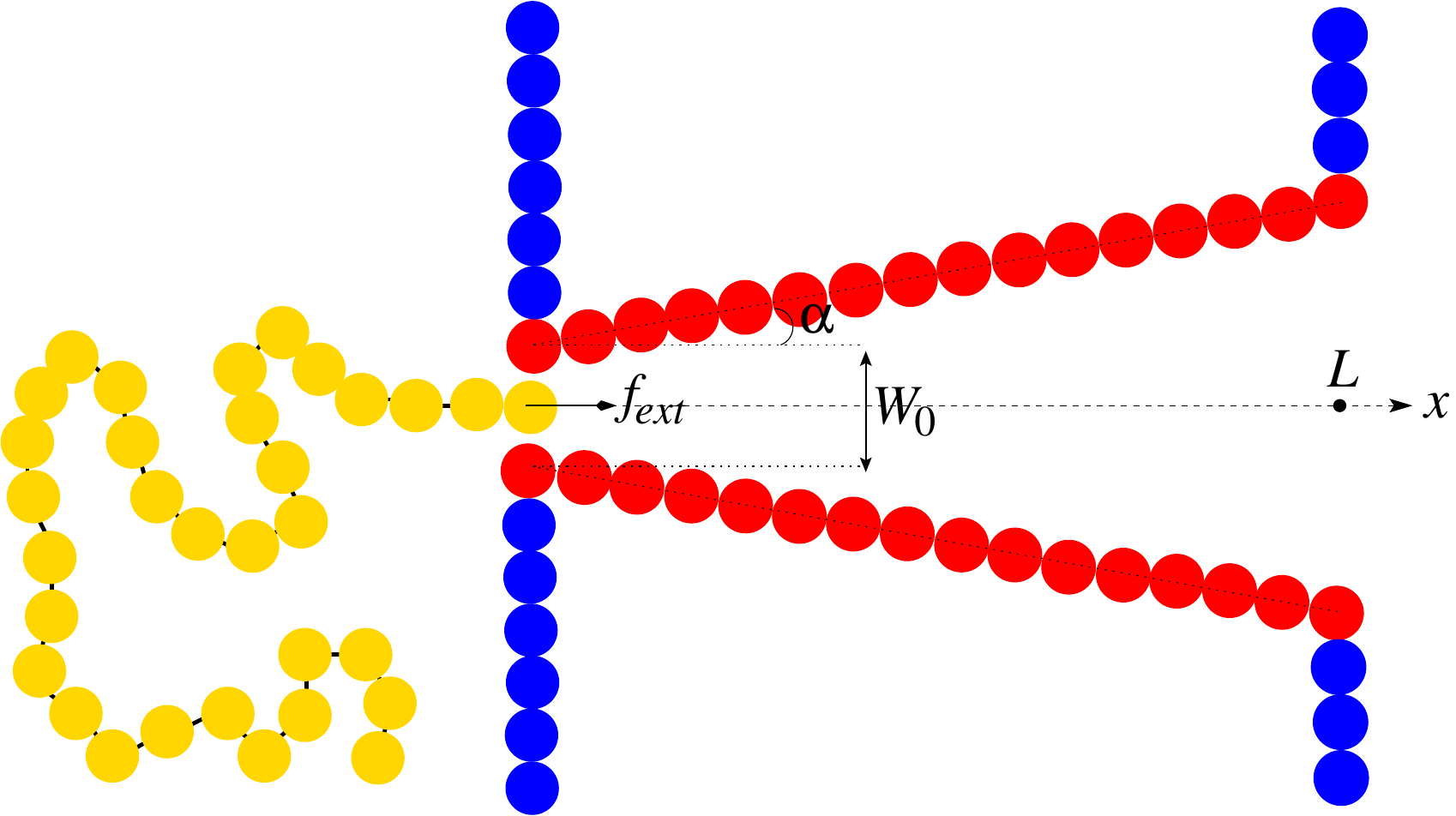}
	
    \caption{ Schematic diagram of a polymer translocating through a
	conical pore with an half apex angle $\alpha$. The width (at the
	apex) and the length of the pore is $W_0=2.25\sigma$ and
	$L=16\sigma$, respectively. \label{fig:1}} 

\end{figure}
}

\newcommand{\figTwo}{
\begin{figure}[t]
\centering
	\includegraphics[width=3.45in]{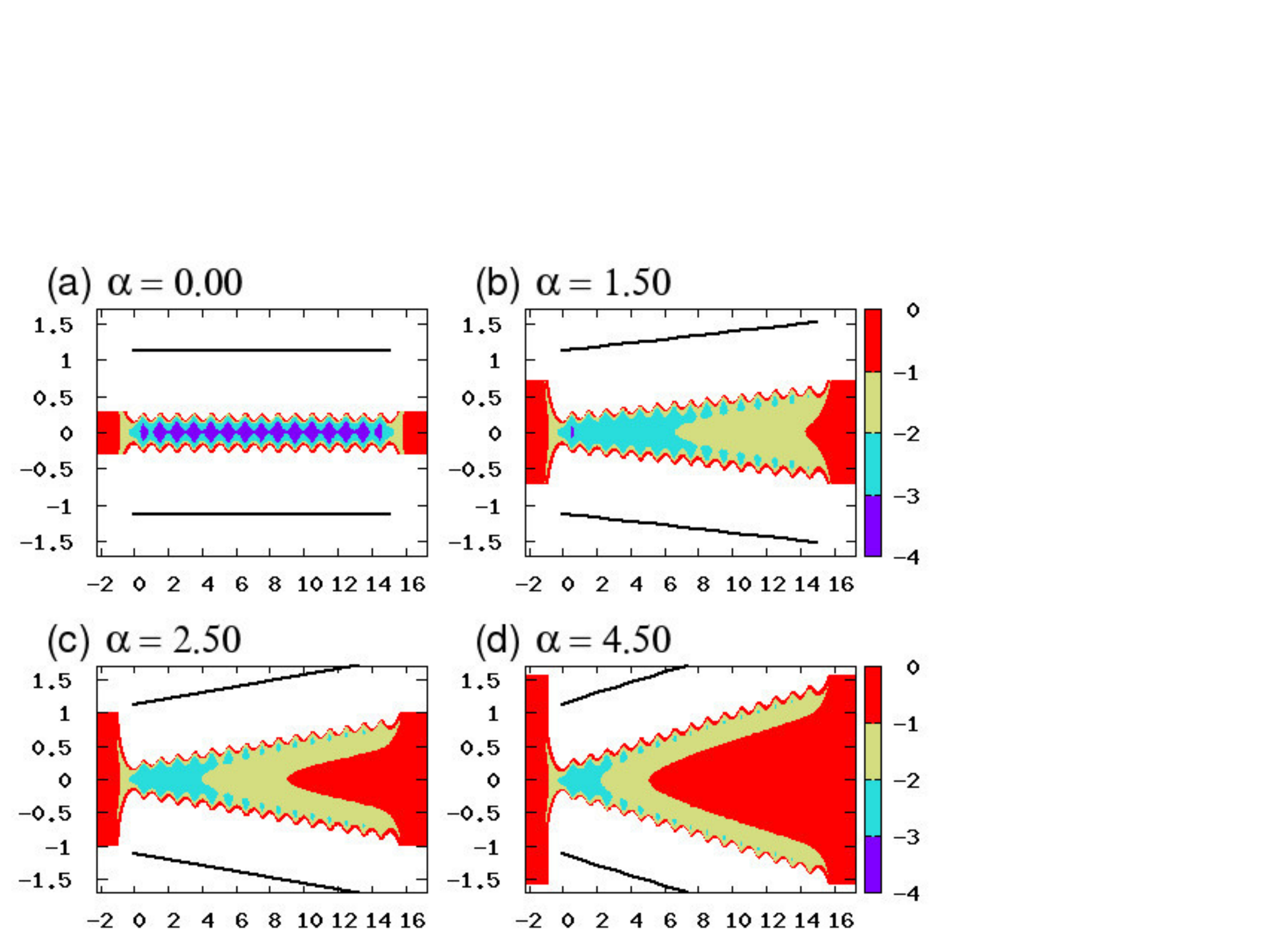}
    
    \caption{\label{fig:2} Potential energy landscape for various pore apex angles $\alpha=0$, 0.25, 1.5 and 4.5 used in our simulations. Potential depth increases from red to purple.}

\end{figure}
}

\newcommand{\figThree}{
	\begin{figure*}[t]
	\centering
  
	\includegraphics[width=0.95\textwidth]{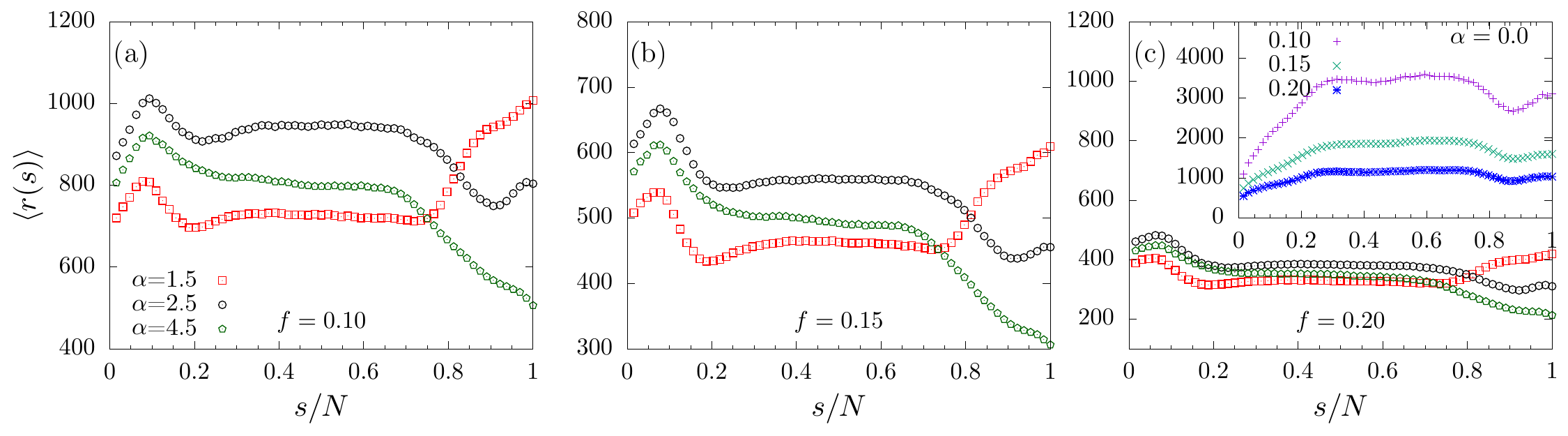}

	\caption{The mean residence time $\langle r(s) \rangle$, for a flexible polymer of length $N=64$ as a function of $s/N$ for various pore half-apex angles $\alpha$ for external driving forces (a) $f=0.1$, (b)  $f=0.15$, and (c) $f=0.2$. The strength of pore-monomer interaction is $\varepsilon_{p}=0.9$. The inset shows the mean residence time for half-apex angle $\alpha=0$ for various values of driving forces.  \label{fig:3}}

\end{figure*}
}

\newcommand{\figFour}{
\begin{figure}[t]
	\centering
	\includegraphics[width=0.35\textwidth]{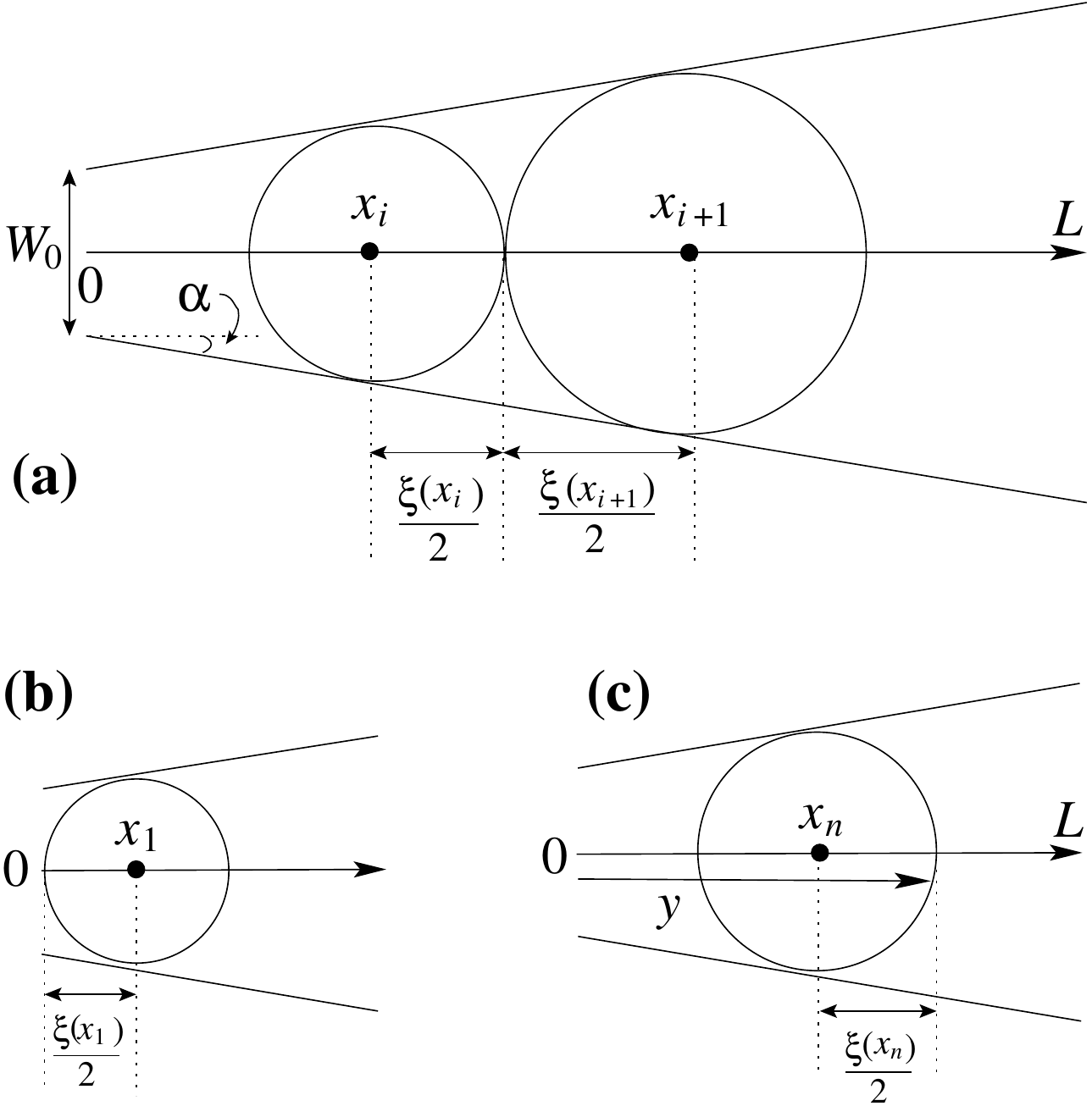}

	\caption{\label{fig:4} (a) Two consecutive blobs inside the conical pore. The diameter $\xi(x_i)$ of a blob depends on the position $x_i$ along the pore axis. This figure also shows the relation satisfied by two consecutive blobs. (b) The location of the first blob is tangent to the beginning of the conical pore. (c) The location of the $n$th blob is at a distance $y$ from the entrance of the conical pore.    
}

\end{figure}
}

\newcommand{\figFive}{
\begin{figure*}[t]
	\centering
	\includegraphics[width=0.98\textwidth]{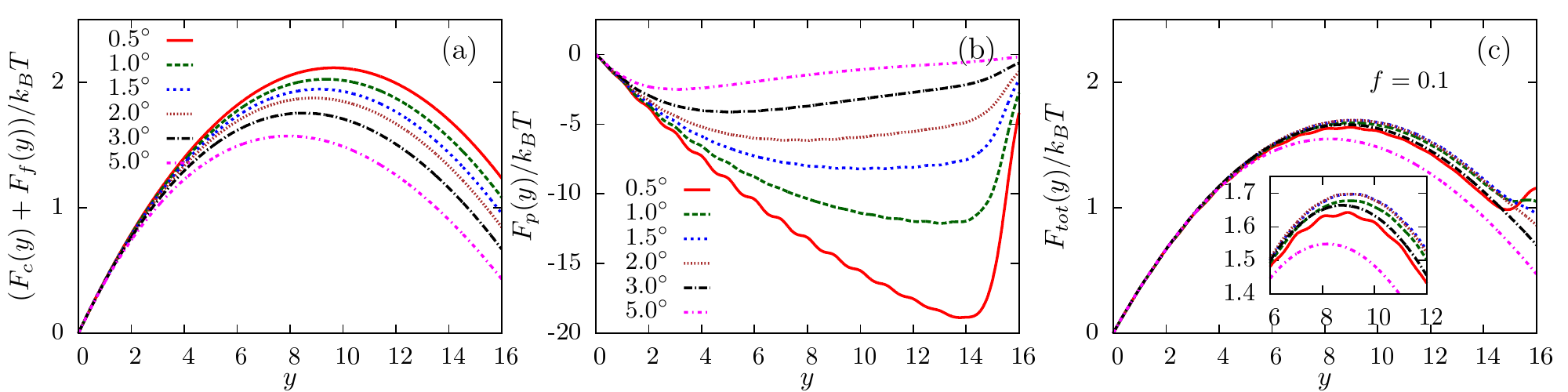}
	
	\caption{(Color online) (a) Sum of free energy contributions due to confinement and external force $(F_{c} + F_{f})/k_BT$, (b) Free energy due to pore interaction $F_{p}/k_BT$, (c) Total free energy $F_{tot}/k_BT$, as a function of distance $y$ from the pore apex along the pore axis for different half-apex angle $\alpha$ for a polymer of length $N=64$ with an external force $f=0.1$. The inset shows the data near the peak region. \label{fig:5} }

\end{figure*}
}

\newcommand{\figSix}{
\begin{figure}[b]
\centering
    \includegraphics[width=3.0in]{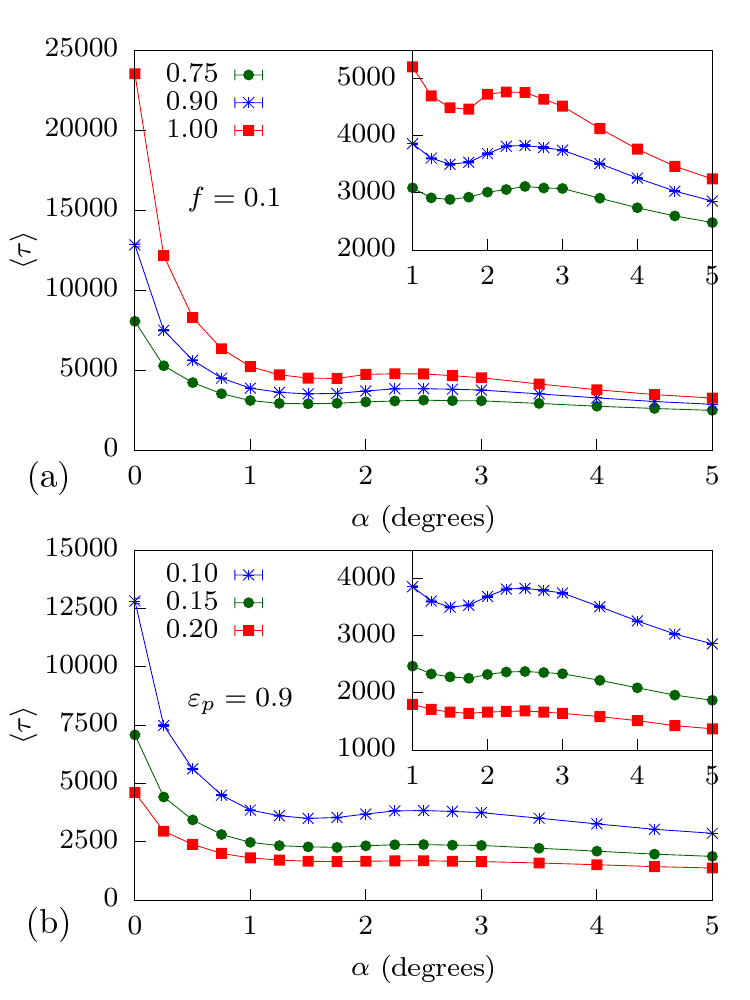}
 
    \caption{\label{fig:6} (a) Average translocation time $\langle \tau \rangle$ as a function of pore apex angle $\alpha$: (a) For various strength of pore-polymer interaction $\varepsilon_{p}$ at pulling force $f= 0.1$. (b) For various pulling forces $f$ by keeping $\varepsilon_{p}=0.9$. The length of the polymer is $N=64$.}  

\end{figure}
}

\def\mr{\langle r(s) \rangle}

\begin{document}

\title{Driven translocation of a flexible polymer through an interacting conical pore}

\author{Rajneesh Kumar} 
\email{rajneesh@jncasr.ac.in, Present address: Theoretical Science
Unit, Jawaharlal Nehru Center for Advanced Scientific Research, Jakkur,
Banglore --560064 India.}

\author{Abhishek Chaudhuri}
\email{abhishek@iisermohali.ac.in}
\author{Rajeev Kapri}
\email{rkapri@iisermohali.ac.in}

\affiliation{Indian Institute of Science Education and Research Mohali,
Knowledge City, Sector 81, S. A. S. Nagar -- 140 306, Punjab India.}



\date{\today}

\begin{abstract}
	
	We study the driven translocation of a flexible polymer through an
	interacting conical pore using Langevin dynamics simulations.  We
	find that, for a fixed value of externally applied force and pore
	polymer interaction strength, the mean residence time of monomers
	inside the pore shows non-monotonic variations with pore apex angle
	$\alpha$.  We explain this behavior using a free energy argument by
	explicitly accounting for pore-polymer interactions and external
	drive. Our theoretical observations are corroborated by the
	simulation results of the mean translocation times as the
	pore-polymer interactions and external driving force are varied.  

\end{abstract}

\maketitle

Polymer translocation is a complex ubiquitous phenomena found in many
biological processes like the passage of mRNA through nuclear pore
complexes, protein translocation across biological membranes through a
narrow channel, injection of DNA from a virus head into a host cell and
gene swapping between
bacteria~\cite{muthukumar2016polymer,palyulin2014polymer}. The
translocation of a polymer through a narrow pore has attracted
considerable attention both
experimentally~\cite{branton2010potential,wanunu2012nanopores,howorka2009nanopore,deamer2016three,merchant2010dna,keyser2011controlling,gu2012detection,venkatesan2011nanopore,meller2003dynamics,movileanu2008squeezing,movileanu2009interrogating,majd2010applications}
and
theoretically~\cite{aksimentiev2010deciphering,milchev2011single,panja2013through,sung1996polymer,lubensky1999driven,muthukumar1999polymer,muthukumar2006simulation,muthukumar2010theory,wong2010polymer,muthukumar2001translocation,kong2004polymer,sakaue2007nonequilibrium,panja2007anomalous,dubbeldam2007driven,gauthier2008monte,gauthier2008monte2,saito2011dynamical,saito2012process,sakaue2016dynamics,ledesma2012length,liu2013ultrashort,rowghanian2011force,ikonen2012unifying,ikonen2012influence,ikonen2013influence,sarabadani2014iso,sarabadani2017driven,sarabadani2018theory,ghosh2020pulling,buyukdagli2019theoretical,cohen2011active,cohen2012stochastic,cohen2012translocation,luo2007influence,luo2008sequence,kumar2018sequencing}
in the last two decades due to its potential in designing inexpensive,
fast and direct sequencing devices for biomolecules like DNA.
Biomolecules such as DNA, moving through a narrow pore blocks the flow
of ionic current through the pore. The resultant ionic current trace is
expected to give information of the DNA sequence that is passing through
the pore. This has led to the extensive use of both biological and
synthetic nanopores as biosensors, for example in sequence detection. 

Biological nanopores such as
$\alpha$-haemolysin~\cite{kasianowicz1996characterization},
MspA~\cite{derrington2010nanopore,butler2008single,manrao2012reading},
phi29~\cite{wendell2009translocation} and recently wild-type aerolysin
nanopore~\cite{cao2016discrimination} have been used to investigate both
single stranded DNA (ssDNA) and double stranded DNA (dsDNA) sequencing.
However, biological nanopores are known to be unstable with changes in
pH, temperature and mechanical oscillations~\cite{howorka2009nanopore}.
Further, the role of pore geometry in the translocation process and the
ability to tune the pore geometry to provide the best sequencing
platform, is difficult to address in the context of biological
nanopores. 
 
 Solid state
 nanopores~\cite{dekker2007solid,wanunu2007chemically,gershow2007recapturing,haque2013solid,luan2012slowing,keyser2006direct},
 on the other hand, have been shown to be stable with respect to
 reasonable changes in pH and temperature and can be fabricated to a
 wide range of shapes and sizes. One of the challenges of nanopore based
 sequencing technologies is that translocation of ssDNA through
 nanopores is extremely fast, the ionic current signatures of single DNA
 nucleotides masked by fluctuations. Therefore sequencing techniques
 have focused at slowing down the transport of ssDNA before readout. In
 this regard, asymmetrical conical nanopores with tapering angles have
 great potential for sensing applications, the constricting geometry
 slowing down translocation of biomolecules and the tip of the cone
 acting like a sensing
 zone~\cite{howorka2009nanopore,keyser2011controlling}. This has led to
 several experimental studies of translocation of biomolecules through
 fabricated solid state conical nanopores~\cite{zhou2017enhanced}.
 Conical nanocapillaries have been used to study detection of DNA and
 other proteins, the conical shape of the pore resulting in ionic
 current rectification at low salt
 concentrations~\cite{harrell2006resistive,lan2011pressure,zhou2017enhanced}.
 An alternative technique using glass nanocapillaries, as opposed to
 solid state nanopores, was capable of simultaneous ionic current and
 fluorescent detection of DNA translocation~\cite{thacker2012studying}.
 Even for biological nanopores such as the $\alpha$-hemolysin, pore
 asymmetry is known to affect the capture rate and current signature of
 DNA depending on its direction of
 transport~\cite{mathe2005orientation}. 
 
 On the theoretical front, there have been only a few studies on polymer
 translocation through conical
 nanpores~\cite{nikoofard2013directed,nikoofard2015flexible,sun2018simulation,tu2018conic}.
 These studies have shown that the shape of the asymmetric conical
 channel gives rise to an effective entropic force that drives
 translocation and that the translocation time is a non-monotonic
 function of the apex angle of the pore. However, in all experimental
 scenarios, the translocation process is driven and the interactions of
 the polymer with the pore is extremely important in determining the
 efficiency of the translocation process. It is therefore imperative to
 understand the extent to which the non-monotonicity in the
 translocation times persists in presence of external drive and varying
 pore-polymer interactions. In
 Refs.~\cite{sun2018simulation,nikolaev2011simulation}, ionic
 conductivity and polymer translocation in the presence of external
 driving in three different conical shaped nanopore geometries were
 studied. Polymer translocation was shown to be heavily dependent on the
 structures of the pore for different pore-polymer interactions and
 external voltage. However, a systematic understanding of the non-linear
 effects of translocation times on the pore-polymer interactions and
 external voltage as the angle of the conical pore is varied, is
 missing. 
 
In this paper, we study the driven translocation of a flexible polymer
through an interacting conical pore with a small entry and a large exit,
using Langevin dynamics simulations. We look at the mean residence time
distribution $\langle r(s) \rangle$ of a monomer $s$ inside the conical
pore. The mean residence time distribution shows non-monotonic features
with the tapering angle $\alpha$ of the pore which are distinctly
different from that through a rectangular channel of same length. We
show how $\langle r(s) \rangle$ can be tuned with $\alpha$ and the
external driving force. We explain the observations using free energy
arguments which includes both the entropic contributions due to
confinement and the synergetics of the pore-polymer interactions.
Finally we show that the non-monotonic behavior of the mean
translocation time with changing apex angle, is strongly affected by
both the pore-polymer interactions and the external driving force.  

\figOne

We model the polymer by beads and springs in two dimensions. A schematic
diagram of a flexible polymer translocating from the \textit{cis} to the
\textit{trans} side of a conical pore is shown in Fig.~\ref{fig:1}. The
beads of the polymer experience an excluded volume interaction modeled
by the Weeks-Chandler-Andersen (WCA) potential of the form
\begin{equation}\label{eq:LJmm}
    U_{\textrm{bead}}(r) = \begin{cases}
        4 \varepsilon \left[ \left( \frac{\sigma}{r} \right)^{12} - \left( \frac{
        \sigma}{r} \right)^{6} \right] + \varepsilon  & \text{for} \ r \le r_{min} \cr
                    0  & \text{for} \ r > r_{min}
    \end{cases}
\end{equation}
where, $\varepsilon$ is the strength of the potential.  The cutoff
distance, $r_{min} = 2^{1/6} \sigma$, is set at the minimum of the
potential. The consecutive monomers in the chain interact via harmonic
potential of the form
\begin{equation}\label{eq:FENE}
    U_{\textrm{bond}} (r) =  K \left( r - r_0 \right)^2,
\end{equation}
where $K$ is the spring constant and $r_0$ is the equilibrium separation
between consecutive monomers of the chain. 

\figTwo
\figThree

The pore and walls are made from stationary monomers separated by a
distance of $\sigma$ from each other. The conical pore is made up of two
rows of monomers symmetric about the $x$-axis with an apex angle $\theta
= 2\alpha$. The length of the pore is taken to be $L=16 \sigma$ with a
width $W_0 = 2.25 \sigma$ at the apex. This pore width allows only
single-file entrance of the polymer and avoid the formation of hairpin
configurations at the apex
opening~\cite{cohen2011active,cohen2012stochastic,cohen2012translocation}.
The interaction of the pore with the polymer is chosen to be the
standard LJ form:
\begin{equation}\label{eq:LJpm}
    U_{\textrm{pore}}(r) = \begin{cases}
			4 \varepsilon_{p} \left[ \left( \frac{\sigma}{r} \right)^{12} - \left(
				\frac{ \sigma}{r} \right)^{6} \right] & \text{for} \ r \le r_{c} \cr
					0  & \text{for} \ r > r_{c},
    \end{cases}
\end{equation}
where $\varepsilon_{p}$ denotes the potential depth and $r_c = 2.5
\sigma$ is the cutoff distance. The interaction of polymer with walls,
$U_{\textrm{wall}}$, is modeled by WCA potential given by
Eq.~\eqref{eq:LJmm}. The potential energy landscape inside the pore for
various pore angles are shown in Fig.~\ref{fig:2}. For small apex
angles, there is a strong attraction near the \textit{cis} side of the
pore and a strong barrier near the \textit{trans} side. As the apex
angle increases, the barrier near the \textit{trans} side shifts toward
the \textit{cis} side of the pore.

The polymer experiences a driving force, ${\boldsymbol f}_{\textrm{ext}}
= f \hat{\boldsymbol x}$ directed along the pore axis with magnitude
$f$, which mimics the electrophoretic driving of biopolymers through
nanopores.  To initiate the translocation process, we start with a chain
configuration with the first bead placed at the entrance of the pore.
This bead is fixed and the remaining beads are allowed to fluctuate so
that the chain can relax to its equilibrium configuration. The first
bead is then released and the translocation of the polymer across the
pore is monitored.  The translocation time $\tau$ is defined as time
elapsed between the entrance of the first bead of the polymer and the
exit of all the beads from the channel. We discard all failed
translocation events.

To integrate the equation of motion for the monomers of the chain we use
Langevin dynamics algorithm with velocity Verlet update. The equation of
motion for a monomer is given by
\begin{equation}
    m \ddot{\boldsymbol r}_i = - {\boldsymbol \nabla} U_{i} +
	{\boldsymbol f}_{ext} - \zeta {\boldsymbol v}_i + {\boldsymbol
	\eta}_i,
\end{equation}
where $m$ is the monomer mass, $ U_i = U_{\textrm{bond}} +
U_{\textrm{bead}} + U_{\textrm{wall}} +  U_{\textrm{pore}}$ is the total
potential experienced by $i$th monomer, $\zeta$ is the friction
coefficient, ${\boldsymbol v}_i$ is the monomer's velocity, and
${\boldsymbol \eta}_i$ is the random force satisfying the
fluctuation-dissipation theorem $\langle {\eta}_i(t) {\eta}_j
(t^{\prime}) \rangle = 2 \zeta k_{B} T \delta_{ij} \delta(t -
t^{\prime})$. The unit of energy, length and mass are set by
$\varepsilon$, $\sigma$, and $m$, respectively, which sets the unit of
time as $\sqrt{m \sigma^2 / \varepsilon}$. In these units, we choose
$\zeta = 1.0$, $K = 10^3 k_BT/\sigma$, $r_0 = 1.12\sigma$ and $k_B T =
1.0$, and the number of polymer beads $N = 64$ in our simulations. A
time step of $\Delta t = 0.005$ is used in all simulation runs.

We first look at the mean residence time, $\langle r(s) \rangle$, of a
monomer $s$ inside the pore. It is defined as the total time spent by
the monomer $s$ inside the pore between its entrance at the pore apex
and its first exit from the base of the conical pore. In
Fig.~\ref{fig:3}, $\langle r(s) \rangle $ is plotted as as a function of
monomer index $s$ for a polymer of length $N=64$ for different values of
pore half-apex angles $\alpha$, and various external driving forces $f$
for a fixed pore-polymer interaction strength, $\epsilon_{p}$.

We first look at $\alpha = 0$, i.e a flat pore. As seen in
Fig.~\ref{fig:3}(c, inset), for small driving forces, $\langle r(s)
\rangle$ increases with $s$ for the initial monomers, then saturates and
finally shows a non-monotonic variation for the end monomers of the
chain. This behavior can be explained using the tension propagation
theory~\cite{sakaue2007nonequilibrium,sakaue2016dynamics,rowghanian2011force,saito2011dynamical,saito2012process,ikonen2012influence,ikonen2012unifying,ikonen2013influence,sarabadani2014iso,sarabadani2017driven,sarabadani2018theory}
with pore-polymer interactions. The part of the polymer on the
\textit{cis}-side is divided into two distinct domains. The external
driving force and the attractive interactions of the pore with the
polymer, pulls on the monomers nearer to the pore and sets them in
motion. The remaining monomers that are farther away from the pore, do
not experience the pull and on average remain at rest. As the polymer
gets sucked inside, more and more monomers on the \textit{cis} side
start responding to the force, with a tension front separating the two
domains propagating along the length of the polymer. The time dependent
drag experienced by a monomer $\zeta(t)$ increases as the tension front
propagates and more number of monomers on the \textit{cis} side get
involved. This increase in the effective friction is manifested in the
residence times which show an initial increase with $s$, implying that
subsequent monomers spends more time inside the pore. This continues
until $\zeta(t)$ becomes maximum when the tension front reaches the last
monomer, maximum number of monomers at the \textit{cis}-side
participating in the translocation process. Beyond this, the system
enters the tail retraction stage where the monomers on the \textit{cis}
side starts decreasing and therefore $\zeta(t)$ decreases and so does
the waiting time $\mr$. For the end monomers exiting the extended pore,
the effect of the attractive interactions of the pore become more
dominant compared to entropic effects which increases their residence
time inside the pore. As the external driving force increases, this
increase in $\langle r(s) \rangle$ for the end monomers is minimal.  

As $\alpha$ increases, we observed several interesting effects which can
be attributed to the conical nature of the pore. For non-zero $\alpha$,
$\mr$ shows a hump for the initial monomers. The initial increase
follows from the argument of tension propagation theory as stated above.
The decrease in $\mr$ observed before the saturation region is a result
of the entropic gain as the initial monomers move into a larger region
than that of a flat pore. The effective drag reduces leading to a
lowering of $\mr$. The saturation regime follows as the tension
propagates to the end monomers. The dominant effect of the pore-polymer
interactions on the end monomers which results in an increase in $\mr$,
reduces as $\alpha$ increases. Further note that $\mr$  is lowest for
all $s$ (except for the end monomers) for $\alpha = 1.5$, increases for
$\alpha = 2.5$ and decreases for $\alpha = 4.5$. This indicates a
non-monotonic dependence of the total translocation time with $\alpha$
consistent with earlier observations. For increasing driving forces, the
subtle competition between entropy and pore synergetics decreases and
the features in $\mr$ are less pronounced.

\figFour
\figFive

We now provide a theoretical description of the translocation process
using a free energy argument. We need to estimate the free-energy change
due to the confinement, $F_c$, of the polymer inside the conical
channel. Let us consider a partly confined chain in the channel in the
presence of an external driving force ${\boldsymbol f}_{ext}$.  The
confinement of the polymer inside the pore costs entropy. We assume that
the part of the polymer chain that is inside the pore breaks up into
blobs of size $\xi (x) = W_0 \cos \alpha + 2x\sin \alpha$ that are
tangent to the pore walls as shown in Fig~\ref{fig:3}. Then, the
entropic penalty due to the confinement of chain in the conical channel
is of the order of $k_B T$ per blob. If $N_{b}(y)$ represents the number
of blobs that penetrate a distance $y$ into the channel, we have
$F_{c}(y) \sim k_B T N_{b}(y)$.

To count $N_{b}(y)$, we consider two consecutive blobs inside the
conical pore at positions $x_j$ and $x_{j+1}$ with diameters $\xi(x_j)$
and $\xi(x_{j+1})$, respectively as shown in Fig.~\ref{fig:4}. They
satisfy the relation 
\begin{equation}
	x_{j+1} =  x_{j} + \frac{1}{2} \left[\xi (x_{j}) + \xi (x_{j+1}) \right].
\end{equation}
Using this recursion relation, one can easily obtain an expression
between the position of the blob $x_j$ and its number $j$ along the
pore~\cite{nikoofard2013directed} as
\begin{equation} 
	\label{eq:xjj}
	x_j = \mathcal{Q}^{j-1} \left( x_1 + \frac{W_0}{2 \tan \alpha}
	\right) - \frac{W_0}{2 \tan\alpha},
\end{equation}
where $\mathcal{Q} = (1+\sin\alpha)/(1-\sin\alpha)$. The first blob is
tangent to the beginning of the pore and its location along the axis is
given by $x_1 = \xi(x_1)/2$. The last (say $n$th) blob is tangent to the
pore at a distance $y$ from the pore entrance, hence its location $x_n$
along the axis is given by $x_n = y - \xi(x_n)/2$. The blob number $n$,
and hence $N_{b}(y)$, can be obtained by Using Eq.~\eqref{eq:xjj} along
with $x_1$ and $x_n$, and hence the free-energy cost due to the
confinement as a function of distance $y$ along the pore axis is given
by
\begin{equation}
	\frac{F_{c} (y)}{k_B T} \sim N_{b}(y) \sim \frac{\log \mathcal{P}} {\log \mathcal{Q}},
\end{equation}
where $\mathcal{P} = 1 + 2y \tan\alpha/W_0$.

The second contribution to the free-energy is due to the constant
external force, $f$, experienced by the segment of the polymer which is
inside the pore, denoted by $F_{f}$. The value of $F_{f}$ changes with
the number of monomers which are present inside the pore.  The free
energy change corresponds to the work done to displace the chain by a
distance $y$ inside the pore and is given by
\begin{eqnarray}
	\frac{F_{f}(y)}{k_B T} &\sim& - \int_0^{y} f N_{b}(x) d x\cr
	&\sim& - \frac{f}{ 2 \log \mathcal{Q} } \left[ 2 y - \left( 2y + \frac{W_0}{\tan \alpha}
\right) \log \mathcal{P} \right],
\end{eqnarray}
where $f$ is the magnitude of the external force ${\boldsymbol f}_{ext}$
acting inside the pore towards the trans direction.

The final contribution to the free-energy is due to the attractive
interactions with the walls of the channel. We need to determine the
number of monomers in the blobs that are in contact with two walls of
the conical channel. In 2D, the number of monomers in $j$th blob is
given by $m(x_j) = \left( \xi(x_j) / \sigma \right)^{4/3}$. The total
number of monomers $N_n$ in $n$ blobs up to the distance $y$ inside the
channel can then be obtained from the constraint $N_n(y) = \sum_{j=1}^n
m(x_j)$. Substituting $x_j$ from the recursion relation
Eq.~\eqref{eq:xjj}, and number of blobs $N_b(y)$, we get
\begin{equation}
	\label{eq:nny}
	N_n(y) = \sigma^{-\frac{4}{3}} \frac{ \left(2y \sin\alpha + W_0 \cos
		\alpha\right)^{\frac{4}{3}} - \left( W_0 \cos\alpha \right)^{\frac{4}{3}} }{ (1+ \sin
		\alpha)^{\frac{4}{3}} - (1-\sin\alpha)^{\frac{4}{3}} }.
\end{equation}
The fraction of monomers that are in contact with a wall of the channel
is then given by $N_n(y)( \sigma / \xi(y))$. If $V(y)$ denotes the
interaction energy, which is due to the pore-polymer interaction and
given by the LJ potential (Eq.~\eqref{eq:LJpm}), the free energy
contribution due to the attractive interaction for a polymer that has
entered the pore up to a distance $y$ is given by
\begin{equation}
	\frac{F_{p}(y) }{k_B T} \sim V(y) N_{n}(y) \left( \frac{\sigma}{ \xi(y)} \right).
\end{equation}
The total free energy is then given by
\begin{equation}
	\frac{F_{tot}(y)}{k_B T} = \frac{1}{k_B T} \left( F_{f}(y) + A F_{p}(y) + C
	F_{c}(y) \right),
	\label{eq:ftot}
\end{equation}
where $A$ and $C$ are undetermined factors. 

In Fig.~\ref{fig:5}, we have plotted the total free energy
$F_{tot}/k_BT$, given by Eq.~\eqref{eq:ftot}, for parameter values
$A=0.035$ and $C=1$, as a function of distance $y$ along the pore axis
for different half-apex angle $\alpha$ ranging from $\alpha=0.5^{\circ}$
to $\alpha = 5^{\circ}$ for a polymer of length $N=64$ with an external
force $f=0.1$ and pore polymer interaction energy strength
$\varepsilon_{p} = 1$. The free energy exhibits a barrier that needs to
be overcome by the polymer to translocate towards the trans side
successfully. As $\alpha$ increases, the barrier height also increases
indicating that it is relatively more difficult for the polymer to
translocate towards the trans end. The free energy barrier attains a
maximum height at $\alpha = 2^{\circ}$ and on increasing $\alpha$
further, the barrier starts decreasing showing that polymer can easily
translocate from the conical pore for larger apex angle. This explains
the non-monotonic dependence of residence times and average
translocation times on the pore angle.

In the above analysis, we have taken the interaction energy, which comes
due to pore-polymer interactions, to be one one-dimensional. However,
the actual interaction energy is two dimensional. The potential energy
landscape presented in Fig.~\ref{fig:2}, clearly indicate that for very
small values of $\alpha$, the position of the potential barrier is
nearer to the trans side of pore, which corroborates the free energy
picture.  So for small apex angles, the translocation process is an
interplay between confinement effects and interactions of the polymer
with the pore.  But as the $\alpha$ increases, position of the potential
barrier start moving towards the apex of the pore as can be seen from
the red region moving close to the pore entry. As the driving force is
increased, the translocation is faster. The polymer will take less time
to overcome the barrier as compared to the smaller values of driving
forces.  At sufficiently strong driving force, effect of pore-polymer
interaction becomes negligible and translocation time is expected to
decrease monotonically with the apex angle of the pore
(Fig.~\ref{fig:2}).

\figSix

In Fig.~\ref{fig:6}(a), we plot the variation of mean translocation time
$\langle \tau\rangle$ as a function of $\alpha$ for different strengths
of pore-polymer interaction, $\varepsilon_{p}$, at a fixed value of the
external force $f=0.1$. As is evident from the free energy argument,
$F_{p}$ dominates the translocation process as $\varepsilon_{p}$
increases. With increasing $\varepsilon_{p}$, the free energy barrier
near the exit of the pore is high and the polymer spends longer time
inside the pore and therefore, $\langle \tau\rangle$ increases with
$\varepsilon_{p}$ for all values of $\alpha$. With increasing $\alpha$
since the barrier falls drastically, the translocation times are
significantly less. However, the non-monotonic feature persists even for
driven translocation with the hump observed near $\alpha \approx
2^{\circ}$. In Fig.~\ref{fig:6}(b), we observe that with increasing
external drive $f$ at a fixed $\varepsilon_{p}$, the mean translocation
time $\langle \tau\rangle$ decreases for all $\alpha$. The non-monotonic
feature starts disappearing at higher $f$. 

We have studied the driven translocation of a flexible polymer through
an interacting cone shaped pore. Conical nanopores have several
advantages in bio-sensing applications and the influence on
translocation behavior of the interactions with the surface of the pore
and the external drive are of significant interest. Using free energy
arguments, we have shown that non-monotonic features in translocation
times observed in conical pores persist even in the presence of the
experimentally relevant scenarios of driven translocation with
pore-polymer interactions. Extensive Langevin dynamics simulations
confirm these observations. The mean residence times also show behaviors
characteristic of translocation through conical pores. Their behavior
with changing external drive is consistent with tension propagation
along the polymer backbone. The mean translocation time $\langle \tau
\rangle$ is found to be strongly dependent on the strength of
pore-polymer interaction and the external driving force.



%

\end{document}